# Plasma Panel Sensors for Particle and Beam Detection

P. S. Friedman, R. Ball, J. R. Beene, Y. Benhammou, E. H. Bentefour, J. W. Chapman, E. Etzion, C. Ferretti, N. Guttman, D. S. Levin, M. B. Moshe, Y. Silver, R. L. Varner, C. Weaverdyck, and B. Zhou

*Abstract*–The plasma panel sensor (PPS) is an inherently digital, high gain, novel variant of micropattern gas detectors inspired by many operational and fabrication principles common to plasma display panels (PDPs). The PPS is comprised of a dense array of small, plasma discharge, gas cells within a hermetically-sealed glass panel, and is assembled from non-reactive, intrinsically radiation-hard materials such as glass substrates, metal electrodes and mostly inert gas mixtures. We are developing the technology to fabricate these devices with very low mass and small thickness, using gas gaps of at least a few hundred micrometers. Our tests with these devices demonstrate a spatial resolution of about 1 mm. We intend to make PPS devices with much smaller cells and the potential for much finer position resolutions. Our PPS tests also show response times of several nanoseconds. We report here our results in detecting betas, cosmic-ray muons, and our first proton beam tests.

## I. INTRODUCTION

THE plasma panel sensor (PPS) was conceived to take advantage of an existing, plasma-TV technology and manufacturing infrastructure for making large area, high definition, plasma display panels (PDPs). PDPs comprise millions of cells per square meter (see Fig. 1), each of which when provided with a signal pulse can initiate and sustain a plasma discharge to illuminate a phosphor. A PPS resembles a PDP, but is modified to detect ionization of the gas in the individual cells. The PPS Geiger-mode discharge is initiated internally by ion-pairs created within the device by an ionizing photon or particle interacting with the detector. The bias voltage across the cell is set to exceed the Paschen potential. The ionizing event creates an electron avalanche (and possibly streamers) that ultimately results in a large gaseous discharge whose amplitude is limited by the cell capacitance. The PPS discharge is terminated by the presence of a localized quench resistance that, combined with the cell capacitance, yields an RC time constant, or cell recovery time long enough that the free charges and metastables in the gas volume are neutralized or deactivated. Depending upon the application, this resistance can be localized at each cell or for each line (as for the prototype tests reported here).

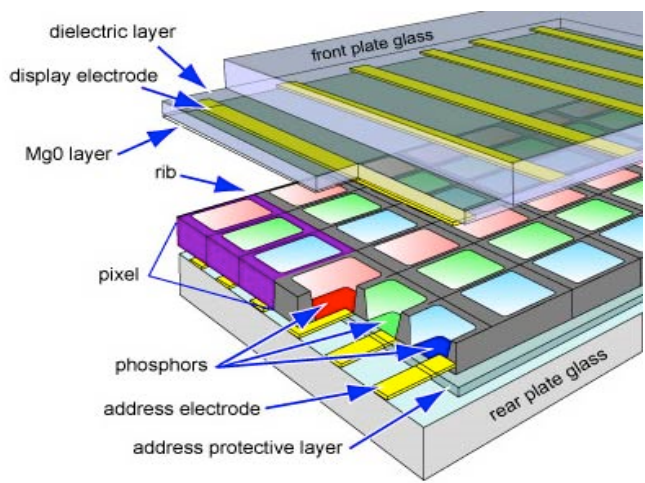

Fig. 1. Typical PDP structure for plasma-TV panel.

Operated this way, the cell configuration and fabrication process is simplified by the elimination of phosphors (i.e. red, green and blue), contrast enhancement and protective layers, rib structures, and thin-film secondary electron emitters (e.g. MgO). Unlike a number of other micropattern gaseous detectors, PPS devices can be hermetically-sealed and are fabricated using stable, non-reactive, inherently radiation-hard materials such as glass substrates, refractory metal electrodes and inert gases.

## II. PPS DEVICE CONFIGURATIONS

A number of PPS device configurations are feasible [1]-[4] with several being investigated, but in all cases each pixel operates like an independent micro-Geiger counter, so the gas discharge can be initiated by either ionization of the gas, or by electrons emitted by a conversion layer in contact with the gas (e.g. for neutron detection) [5]. Our focus however, has been primarily on tests using PPS devices fabricated from modified PDPs. These devices are able to detect charged particles by direct gas ionization [6].

We show in Fig. 2 a columnar-discharge PPS with an open-cell orthogonal X-Y electrode structure. By "open-cell" we mean that there is no rib enclosure surrounding each cell as shown in Fig. 1 for PDP TVs. The discharge occurs in the volume defined by the intersection of the front column

Manuscript received November 17, 2012. This work was supported in part by the U.S. Department of Energy under Grant Numbers: DE-FG02-07ER84749, DE-SC0006204, and DE-SC0006219. This work was also partially supported by the Office of Nuclear Physics at the U.S. Department of Energy, and by the United States–Israel Binational Science Foundation under Grant No. 2008123.

P. S. Friedman is with Integrated Sensors, LLC, Ottawa Hills, OH 43606 USA (telephone: 419-536-3212, e-mail: peter@isensors.net).
R. Ball, J. W. Chapman, C. Ferretti, D. S. Levin, C. Weaverdyck and B. Zhou are with the Department of Physics, University of Michigan, Ann Arbor, MI 48109 USA.
J. R. Beene and R. L. Varner are with the Physics Division, Oak Ridge National Laboratory, Oak Ridge, TN 37831 USA.
Y. Benhammou, M. B. Moshe, E. Etzion, N. Guttman and Y. Silver are with the School of Physics and Astronomy, Tel Aviv University, Tel Aviv 69978 Israel.
E. H. Bentefour is with Ion Beam Applications S. A., Louvain La Neuve, B-1348 Belgium.

electrodes (e.g. HV-cathodes) and the back row electrodes (e.g. sense anodes) as shown in Fig. 2. The discharge/gas gap can typically vary from a few hundred micrometers to a few millimeters, depending upon the application. The electrode width will similarly vary over approximately the same range.

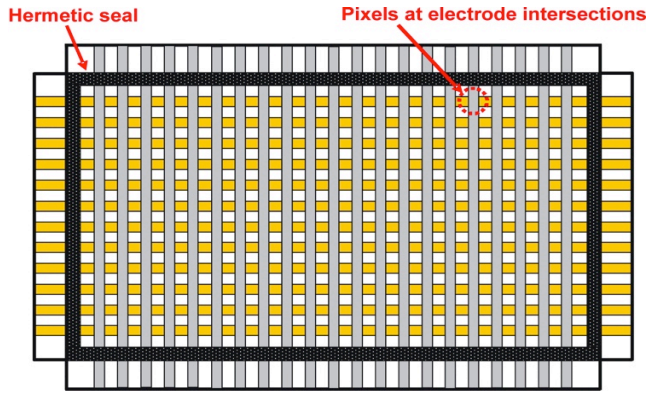

Fig. 2. Columnar-discharge PPS electrode structure.

Fig. 3 shows a columnar-discharge PPS test panel having the orthogonal electrode structure in Fig. 2, after modifying a commercial 2-electrode, DC-type, glass PDP. The panel in Fig. 3 is attached to a removable aluminum frame for mechanical integrity, which is fitted with a sealed, high-vacuum, shut-off valve to allow multiple fills of different gas mixtures and pressures. The panel active area is 8.1 cm × 32.5 cm, and we have used both transparent $SnO_2$ and Ni column HV-electrodes (i.e. cathodes), and Ni back row sense anodes. The electrode pitch is 2.5 mm. These panels, initially designed as monochromatic displays, undergo a systematic bake-out and gas fill procedure before being operated as detectors. They have produced the gas discharge pulses and data reported in this paper. In this configuration, with small gaps (~ 400 μm) relative to the 1.4 mm electrode widths, the field between anode and cathode is fairly uniform, as determined by a COMSOL modeling [1]. A readout electronics card mounts on the horizontal anode lines and the signal is picked off using a 50 ohm termination resistance. A high voltage bus feeds the vertical cathode lines via a single quench resistance per line.

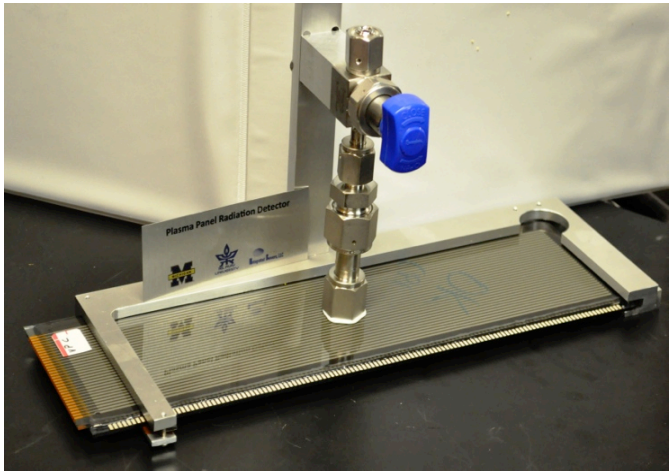

Fig. 3. PDP "refillable" test panel.

The refillable PPS test panel in Fig. 3 has proven more durable than initially expected, as we can typically hold a given gas mixture for months without observing any change in performance. In fact our best test panel to date is still operational more than eight months after the shut-off valve was closed! By being able to use the same panel with different gas mixtures, we can study the effect of gas composition and pressure completely isolated from any uncertainty associated with panel-to-panel variations in: discharge and/or gas gap, electrode line width, thickness and surface condition, substrate thickness and dielectric surface variation, etc. We are now in the process of modifying similarly constructed panels to Fig. 3, but with a pixel pitch of 1.0 mm and 0.6 mm. These panels however, have not yet been coupled to our readout electronics.

III. EXPERIMENTAL SETUP

We have constructed two test benches, one at the University of Michigan (U-M) and the other at Tel Aviv University (TAU). Each test bench includes a gas delivery system, a triggering system, and a data acquisition (DAQ) system. At these labs, we use beta-emitters, Sr-90 and Ru-106, and cosmic-ray muons as our test radiation. We also have access to a ProCure medical proton beam accelerator near Chicago through an informal collaboration with Belgium proton beam therapy manufacturer Ion Beam Applications S.A. (IBA). We used their Model C235 accelerator to test our devices with a 226 MeV collimated proton beam using aperture diameters of both 1 mm and 10 mm. The triggering system for our lab based experiments is done with a scintillator hodoscope (see Fig. 4), or relies on self-triggering. The proton test beam data were acquired with a PPS self-trigger.

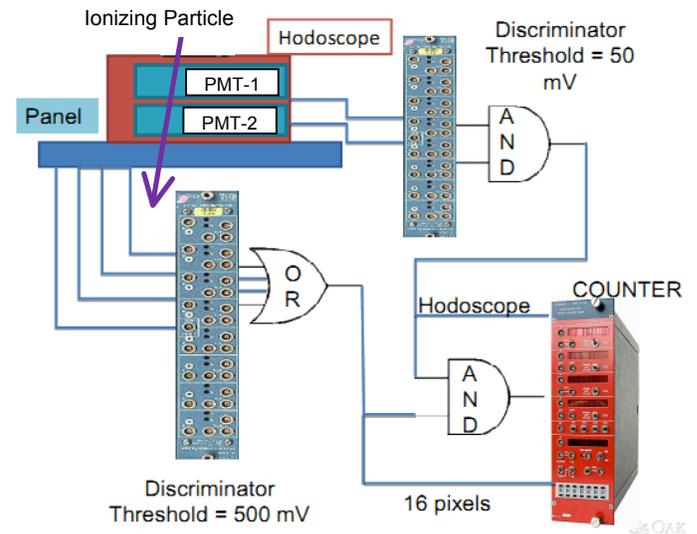

Fig. 4. Hodoscope coincidence measurement setup.

Our current DAQ system is adapted from the Muon Spectrometer monitored drift tube (MDT) readout electronics developed (in part by U-M) for the ATLAS experiment at the Large Hadron Collider (LHC). The first generation of the new DAQ readout electronics has the capability to acquire data for 24 channels with nanosecond resolution.

## IV. RESULTS AND DISCUSSION

We have investigated the PPS device response to a number of ionizing particle sources under different experimental conditions with various discharge gases. The discharge gases tested include: $Ar+CO_2$, $Ar+CF_4$, $CF_4$, $SF_6$ and Xe. For a few of them the pressures have ranged from about 200 to 700 torr, but here we report results at a single pressure of 600 torr. The observed signals from all of the devices tested have had large amplitudes of at least several volts, so there has been a need for attenuation instead of amplification electronics. For each gas tested the shape of the induced signals is uniform. The leading edge rise time for the current generation of panels is typically 1 to 2 ns (see Fig. 5). Not unexpectedly the device performance has been shown to be very much gas dependent, with the operating voltages varying by more than 1000 volts for different gas mixtures in the same panel.

For the experiments reported here, we have employed four different particle sources: betas from $^{90}$Sr (max. electron energy of 2.3 MeV), higher energy betas from $^{106}$Ru (max. energy of 3.5 MeV), relativistic particles/energies from cosmic muons ($\geq$ GeV), and 226 MeV protons from an IBA-C235 accelerator. In all cases the actual signal pulses appear similar (see Fig. 5) for a given panel geometry, gas mixture, cathode voltage, and quench and signal resistors. In other words, the signal amplitude, rise time and duration do not appear to depend on the event causing the initial gas ionization. There is nothing surprising about this observation as the cells are being driven in the Geiger or gas breakdown mode.

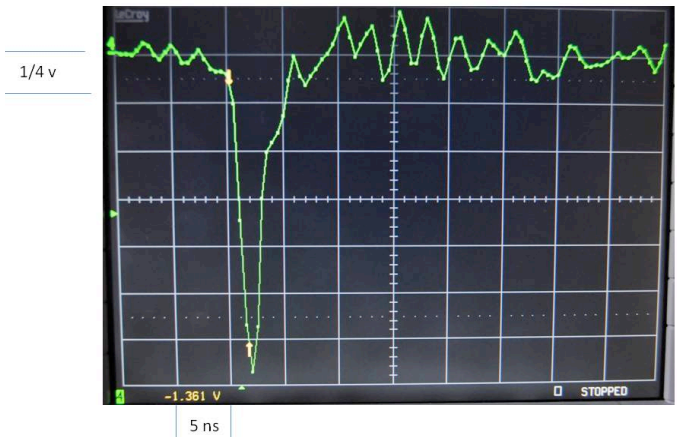

Fig. 5. Typical signal pulse for columnar-discharge PPS.

A typical PPS gas discharge pulse is shown in Fig. 5 from a panel similar to that in Fig. 3, filled with 1% $CO_2$ in 99% Ar at 600 torr, operated at 840V. The experiment employed a $^{106}$Ru beta-source in conjunction with a two-fold coincidence hodoscope (i.e. trigger) as in Fig. 4. The rise time was ~ 1 ns with a 1.9 ns pulse duration (FWHM). Depending on the specific gas and discharge high voltage, the signal amplitudes can range from a few volts to many tens of volts. These large amplitudes result from the effective discharge capacitance for these PPS panels including contributions from neighboring electrodes.

For a given panel and gas mixture, we can generate a PPS *characteristic response curve* of dependence of the rate on the HV quench resistance, as shown in Fig. 6. The panel response is the rate of hits detected and is plotted as a function of the reciprocal of the line quench resistor.

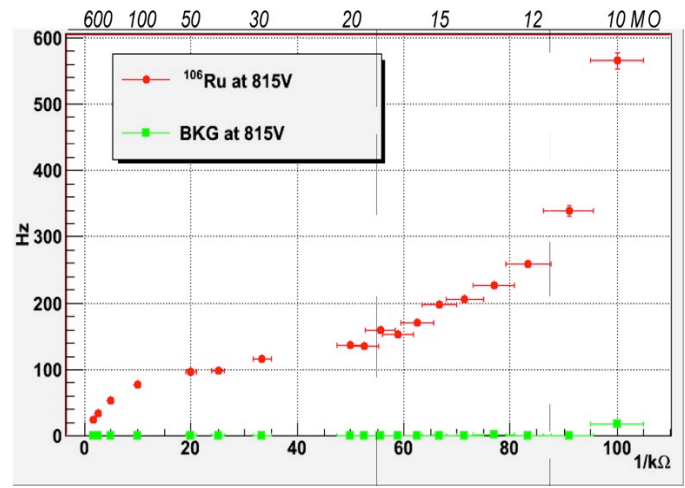

Fig. 6. PPS characteristic response curve: PPS response vs 1/quench resistance

In order to be representative of the panel, the data of this curve are the response sum over several cells on a given HV-cathode line so as to be indicative of the average panel performance for a given line quench resistor (ideally the curve should also be generated for more than one HV-line). For the data shown in Fig. 6, the panel gas was 1% $CO_2$ in Ar at 600 torr (i.e. same as Fig. 5) and was operated at 815 V. The radiation source was $^{106}$Ru and the hits were collected on a single HV line (#110), across four readout lines (RO = 3-6). The quench resistors covered the range from 10 to 600 M$\Omega$.

As suggested by Fig. 6, the PPS characteristic response curve can be analyzed as consisting of three different response regions. For very high quench resistance values, 100 to 600 M$\Omega$, the PPS response rate drops quickly because a high RC time constant means that each pixel is dead for a longer time and the maximum line rate is limited by the HV recovery frequency (order of magnitude ~ 1/RC). At the other end of the curve, 10 to 25 M$\Omega$, the PPS response rate increases quickly as the quench resistance drops. This is caused by a small RC time constant that allows the HV to return to discharge potential before all of the charged species in the cell can be neutralized. This, in turn, leads to after-pulses due to regeneration resulting in *artificially* high count rates. Another contributor of equal or greater importance to regeneration is gaseous metastable species that also have not yet had enough time to decay. Finally the most important region in terms of device optimization is the semi-flat region defined by the moderate quench resistance values between about 25 to 100 M$\Omega$. This is the range of "moderate" quench resistance values and moderate RC time constants, in which we see minimal rate dependence on the quench resistor value. For the panel in Fig. 6, the response rate in this region is ~ 100 Hz.

Another significant result illustrated in Fig. 6 is the PPS response with no source present. The measured background rate is minimal across the entire quench resistance region. This behavior is similar to the very low background rates observed over a large range of signal producing voltages that we reported previously for a panel with transparent $SnO_2$ cathodes and filled with $CF_4$ at 500 torr [6]. In general, PPS devices appear to have low background counts. Although low background count rates in the absence of an efficiency measurement can be misleading, we consider the measured low rates to be a promising indication of good performance.

In addition to the low background rates discussed above, we have previously shown for panels such as in Fig. 3, filled with fluorinated discharge gases, that the arrival time jitter ($\sigma$) as measured using cosmic-ray muons is $\leq 5$ ns [1], [6].

A critically important PPS parameter for most applications is the device position resolution. We measured it acquiring data while translating a "collimated" $^{106}$Ru beta-source through a 1.25 mm wide graphite slit (20 mm thick) in 0.5 mm increments across the sense electrodes in the PPS shown in Fig. 3, filled with the same 1% $CO_2$ in Ar gas mixture as in Fig. 5, but at 890 volts. The plot in Fig. 7 shows the Gaussian means of the hit distribution over the 24 channels readout vs. the source position. The RMS position resolution is 0.7 mm, in a panel with a 2.5 mm electrode pitch. We obtain a slope of $0.39 \pm 0.01$ mm$^{-1}$, where the error is estimated from fitting the plot over three ranges, a value consistent with the electrode pitch.

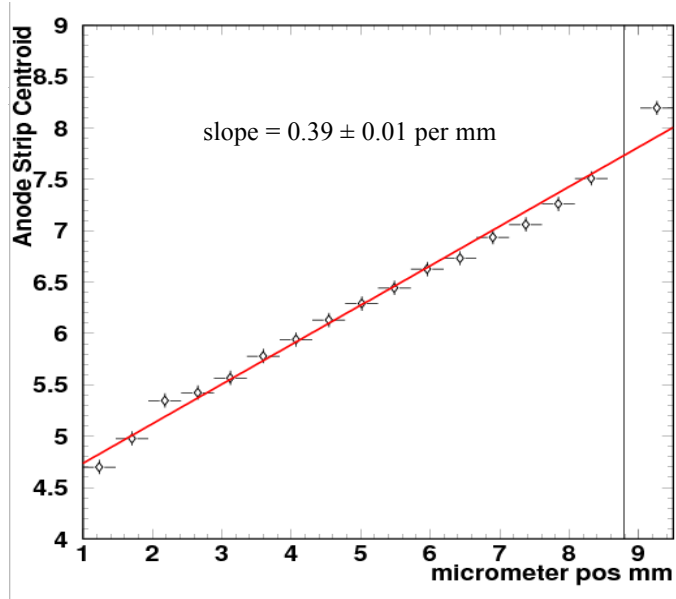

Fig. 7. Beta-scan position resolution measurements.

In order to evaluate the contribution to the position resolution of the spreading/scattering of source emitted electrons, we simulated the measurement with GEANT4. The incoming electrons were described by a pencil beam of beta particles emanating out of the $^{106}$Ru source and traveling through the 20 mm long air gap of the 1.25 mm wide graphite collimator and then through the 2.25 mm thick glass substrates of the PPS. The simulation also included the scattering contribution of the beta particles through the 0.44 mm path length of Ar discharge gas at 760 torr. A total of 1,000,000 tracks were run for the GEANT4 simulation, with a computer generated representation of a sub-sample of 1,000 random tracks shown in Fig. 8. As can be seen in Fig. 8, most of the scattering and absorption of betas occur in the PPS front glass substrate with very few betas exiting the back glass substrate.

This is why our coincidence experiments could not be easily performed using the lower energy $^{90}$Sr beta-source. Even with higher energy betas from the $^{106}$Ru source, significant time is required to accumulate a statistically reproducible distribution. This is one reason why relativistic muons and accelerated protons are so useful for this type of experiment, as the much higher energy of these particles is more than sufficient to penetrate the scintillator and glass layers, although for cosmic-ray muons the time required is very long due to their low intensity.

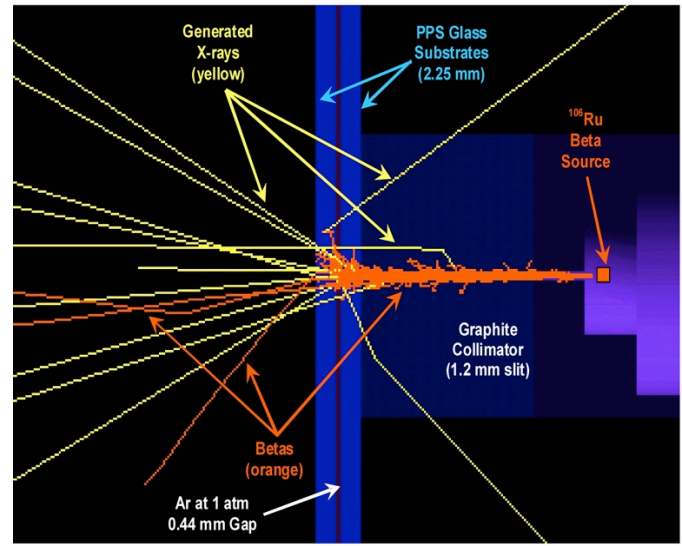

Fig. 8. GEANT4 beta scattering simulation with $^{106}$Ru source.

From the above numerical simulation analysis, we see that the initial 1.25 mm collimated beam of beta particles has a scattering radius of about 5 mm or two lines by the time it reaches the discharge gas. In other words, the "collimated" beta beam inside the PPS scatters approximately two adjacent sense electrodes on each side of the targeted electrode under the graphite slit. Given this incident particle dispersion, the fact that we are able to resolve the beam centroid to within almost a quarter of a pixel (i.e. 0.7 mm in a PPS with a cell pitch of 2.5 mm) bodes very well for the potential position resolution of these devices. In this regard we are currently in the process of fabricating next-generation PPS devices with a cover plate thickness of 0.38 mm (compared to the current 2.25 mm thickness), and eventually plan to fabricate such devices with an electrode pitch of ~ 0.15 mm. We expect that such PPS devices should have a position resolution of $\leq 50$ μm.

We performed our first particle beam experiments with an IBA-C235 proton beam accelerator used for proton therapy (i.e. treatment of cancer). In Fig. 9(Right) we show the number of hits per channel during a position scan using an intense (i.e. > MHz) 1 mm diameter, 226 MeV proton beam for 16 sequential runs in which the panel in Fig. 3 was shifted in each run by ~ 1 mm increments relative to the fixed position proton beam. Each bin is a single data channel for a sense-electrode line. Fig. 9(Left) shows the reconstructed position centroid of the "hit" map from Fig. 9(Right) versus the PPS relative displacement in millimeters with respect to the initial position. The position centroid for each run is based on the weighted average over 3 bins around the peak, approximately matching the 2.5 mm electrode pitch. As with the beta position resolution scan in Fig. 7, the resulting slope of the linear fit (p1 in the legend) establishes that the panel was able to reproduce the proton beam position.

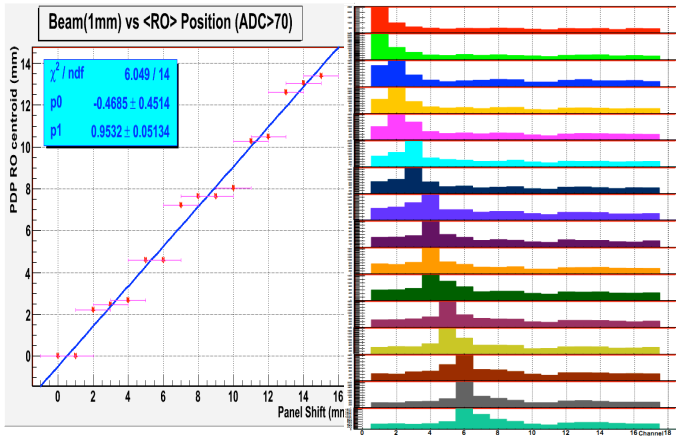

Fig. 9. Position scan measurements with intense 1 mm diameter proton beam.

The steps observed in the Fig. 9(Left) data are presumed to be caused by the intense beam saturating the central pixels. This saturation derives from the deliberately long time constants chosen for this first proton beam test.

To further look into PPS saturation we investigated the response to the simultaneous exposure to two sources in an experiment as follows: four adjacent 32 cm long signal readout (RO) lines (i.e. sense row electrodes) were connected to discriminators whose outputs were OR'ed and then their combined signal rates were measured with a rate counter. High voltage (HV) was applied to two transverse column electrodes (i.e. cathodes) at varying distances from one another. Specifically, HV was applied always to one fixed line (#110) while the second line receiving high voltage was allowed to vary from #100 up to #110 (see Fig. 10).

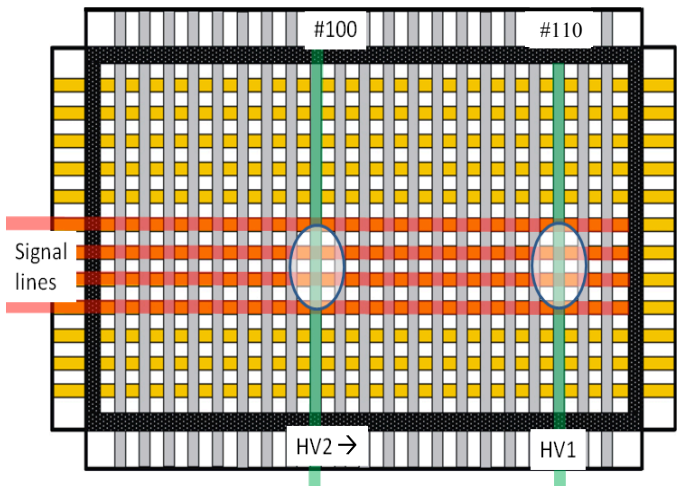

Fig. 10. Configuration for double source test. Shaded regions show approximate location of radioactive beta sources. The line labeled HV2 is incremented from left to right towards HV1.

The intersections of the isolated HV electrodes with the four readout electrodes constituted the active pixels in this test. Each set of four pixels was exposed at first separately, and then simultaneously to two *partially* collimated sources ($^{90}$Sr and $^{106}$Ru) yielding approximately similar rates of betas entering the gas gap region. These sources were positioned, one below the panel and one above, over the active pixels as indicated by the two oval shaded regions in Fig. 10. The second source position was incremented from left to right across the panel starting from line #100. As in the proton beam test, a large quench resistance was deliberately selected in order to produce long cell recovery times close to the saturation value along the high voltage line. The rates of the two groups of pixels were measured when exposed independently and then simultaneously to the two sources.

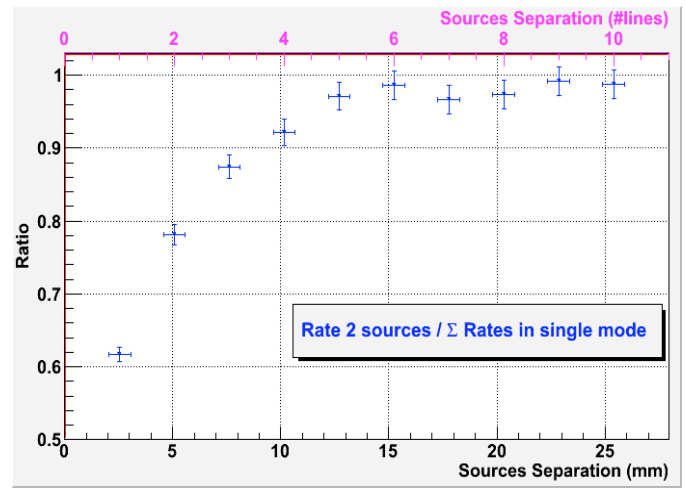

Fig. 11. Ratio of the rate from two simultaneous sources divided by the sum of the two rates from the same sources separately.

The rate of the four RO lines measured when both sources are simultaneously used equals the linear sum of the rates from two sources when measured individually over nearly the entire width of the panel, which results in the near unity ratio across most of Fig 11. Significant deviations are observed when the two sources are brought within a few lines of each other; in particular when their separation falls below 10 mm - i.e. within 4 lines. As discussed previously, each source has a scattering radius in the PPS of about 5 mm (actually the dispersion is slightly worse than in the GEANT4 simulation for Fig. 8 because the two sources were only *partially* collimated). From Fig. 11 we observe that starting from a separation distance of 7.5 mm (i.e. line #107 in Fig. 10), the double source rate decreases below 90% of the sum of the two rates in single mode. When a single source is used over a HV line, the total rate is increased by betas scattered over the other HV line if it is close enough. But when both sources are used at the same time, and both HV lines are respectively saturated by their corresponding source, then when both sources are close enough to overlap in terms of their scattering radius (e.g. with sources on lines #107 and #110) the rate increment due to the overlapping scattered electrons cannot happen. Hence the reduced ratio observed in Fig. 11 (i.e. starting at line #106 and dropping below 90% for line #107). The initial experimental results of the double radiation source tests indicate that the saturation effect is quite limited in extent. Our new generation of PPS structures are being fabricated with a 0.38 mm cover plate thickness and should result in much less scattering of incident beta radiation, as well as less capacitive coupling and reduced saturation, and should thus allow us to further improve the resolution of adjacent cell hits by separate sources.

## V. CONCLUSIONS

A few properties of our current PPS devices are described in this paper with data from PDP commercial panels modified to function as ionizing radiation counters. Like PDPs, our PPS devices are inherently digital, low noise and have high gain. They can also be hermetically sealed, thus eliminating the complexity associated with a number of other micropattern gaseous detectors that require a continuous gas flow support system. However, even without a *hermetic* seal, we have developed a mechanical valve/seal technology together with a panel baking and gas filling procedure that allows each panel to operate as a stable, portable test chamber for evaluating the PPS device performance as a function of the discharge gas mixture and pressure. The measurement of a PPS characteristic response curve of a panel (depending on its structure, on the gas mixture and on the HV applied), allows one to select a quench resistance value to work in a region where the hit rate is weakly dependent on the external HV resistance. This is a first important step toward a good evaluation of the efficiency of the PPS.

We have demonstrated particle detection for betas, protons and cosmic-ray muons, with pulse rise times of 1 ns, pulse widths (FWHM) of 2 ns, and a temporal response or timing jitter of 5 ns [1]-[6]. Dedicated experiments show a remarkable position resolution, much better than the pixel pitch even at the test beam with an unfavorable configuration and lines in saturation. A specific experiment with two sources was performed to measure the saturation effect, and the results are in very good agreement with the results of our GEANT simulation of the source particles scattering.

To complement our experimental program, we have instituted a modeling and simulation effort that has already proved useful based on a toolkit primarily involving GEANT4, COMSOL and SPICE [1]. In the future we will use these tools to investigate the performance advantages and limitations of new PPS designs for specific applications as well as for device optimization.

In summary, we have demonstrated device sensitivity to independent and separate high-intensity radiation sources. We have shown that for a given panel structure and gas, the discharge signals look remarkably uniform regardless of the source of ionizing particles. As we transition to smaller cell sizes with better cell physical and electrical isolation, we expect to achieve lower capacitance and faster discharge times in the sub-nanosecond range, very high position resolution, and excellent response to high luminosity sources. We believe that the fast rise times and short pulse durations are largely due to the very high gain of the Geiger-mode electron avalanche.